\documentclass
[aps,preprint,onecolumn,showpacs,showkeys,10pt,a4paper,floats]{revtex4}%
\usepackage{amsfonts}
\usepackage{amsmath}
\usepackage{amssymb}
\usepackage[final,dvips]{graphicx}
\usepackage[dvips]{rotating}
\usepackage{hyperref}%
\ifx\pdfoutput\relax\let\pdfoutput=\undefined\fi
\newcount\msipdfoutput
\ifx\pdfoutput\undefined\else
\ifcase\pdfoutput\else
\ifx\paperwidth\undefined\else
\ifdim\paperheight=0pt\relax\else\pdfpageheight\paperheight\fi
\ifdim\paperwidth=0pt\relax\else\pdfpagewidth\paperwidth\fi
\fi\fi\fi
\begin{document}
\title{The Inglis-Belyaev formula and the hypothesis of the two-quasiparticle excitations}
\author{B. Mohammed-Azizi}
\email{aziziyoucef@voila.fr}
\affiliation{University of Bechar, Bechar, Algeria}

\begin{abstract}
The goal of the present work is to revisit the cranking formula of the
vibrational parameters, especially its well known drawbacks. The latter can be
summarized as spurious resonances or singularities in the behavior of the mass
parameters in the limit of unpaired systems. It is found that these problems
are simply induced by the presence of two derivatives in the formula. In
effect, this formula is based on the hypothesis of contributions of excited
states due only to two quasiparticles. But it turns out that this is not the
case for the derivatives. We deduce therefore that the derivatives are not
well founded in the formula. We propose then simply to suppress these terms
from the formula. Although this solution seems to be simplistic, it solves
definitively all its inherent problems.

\end{abstract}
\endpage{ }
\volumeyear{ }
\maketitle

\section{Introduction}

Collective low lying levels of the nucleus are often deduced numerically from
the Interacting Bosons Model (IBM) \cite{01} or the Generalized Bohr
Hamiltonian (GBH) \cite{1}-\cite{1a}. Restricting ourselves to the latter we
can say that it is built on the basis of seven functions: The collective
potential energy of deformation of the nucleus, and for its kinetic-energy
part, three mass parameters (also called vibrational parameters ) and three
moments of inertia. All these functions depend on the deformation of the
nuclear surface. Usually, the deformation energy can be evaluated in the
framework of the constrained Hartree-Fock theory (CHF) or by the
phenomenological shell correction method. The mass parameters and the moments
of inertia are often approximated by the cranking formula \cite{2}-\cite{2a}
or in the self consistent approaches by other models \cite{3}-\cite{5}. Most
of the self-consistent formulations are based on the adiabatic time dependent
Hartree-Fock-Bogolyoubov approximation (ATDHFB) which leads to constrained
Hartree-Fock-Bogolioubov (CHFB) calculations \cite{5a}-\cite{5b} in which the
so-called Thouless-Valatin corrections are neglected. It is to be noted that
there are several self-consistent formulations for the mass parameters in
which always some approximations are made (not always the same). Other types
of approaches of the mass parameters use the so-called Generator Coordinate
Method combined with the Gaussian-Overlap-Approximation (GCM+GOA) \cite{5b1}.
Recently new methods have again been developped \cite{5bb}-\cite{5b2}. This
leads to a certain confusion and the problem of the evaluation of the mass
parameters remains (up to now) a controversial question as already noticed in
Ref.\cite{5b2}.

In this paper we will focus exclusively on the mass parameters, especially on
the problems induced by the cranking formula, i.e. the "classical"
Inglis-Belyaev formula of the vibrational parameters. Indeed, it is well-known
that this formula leads sometimes to inextricable problems when the pairing
correlations are taken into account (by means of the BCS model). The
transition between normal $\left(  \Delta=0\right)  $ and superfluid phase
$\left(  \Delta>\Delta_{0}\approx0.3MeV\right)  $ affects generally the magic
nuclei near the spherical shape under the changing of the deformation
\cite{11}. The problem occurs sometimes (not always) exactly in these cases
for an unpaired system $\Delta\sim0.$ In that cases the mass parameters take
anomalous very large values near a "critical" deformation close to the
spherical shape.

This singular behaviour is well-known and constitutes undoubtedly unphysical
effect. It has been early found that these problems are due simply to the
presence of the derivatives of $\Delta$ (pairing gap) and $\lambda$ (Fermi
level) in the formula. They have been reported many times \cite{1},
\cite{10}-\cite{14} in the litterature, but no solution has been proposed. The
authors of Ref. \cite{1} and \cite{11} claim that for sufficiently large
pairing gaps $\Delta$ the total mass parameter is essentially given by the
diagonal part without the derivatives, whereas those of Ref. \cite{14} affirm
that the role of the derivatives is by no mean small in the fission process
and this leads to contradictory conclusions. Other studies \cite{13} neglect
the derivatives without any justification. Some self-consistent calculations
met also the same difficulties. For example in Ref. \cite{5c}, resonances in
mass parameters have also already been noticed. As in the present work they
were attributed to the derivative of the gap parameter $\Delta$ near the
pairing phase transition. In short, up to now the problem remains unclear.
\newline Curiously, one must point out that contrary to the vibrational
parameters, the same formulation (Inglis-Belyaev) for the moments of inertia
does not exhibit any explicit dependence on $\Delta$ and $\lambda$ (as the I-B
formula does for the mass parameters) and this explains why the I-B formula
for the moments of inertia does not meet such problems. This difference
appears not so natural and is a part of the motivation of this work. All these
problems as well as intensive numerical calculations led us to ask ourselves
if the presence of these derivatives is well founded. If this is not the case,
their removal should be justified. In fact, the Inglis-Belyaev formula is
based on the fundamental hypothesis on contributions of two-quasiparticle
states excitations. Rigorously it turns out that the derivatives of $\Delta$
and $\lambda$ do not belong to this type of excitations and this must explain
their reject from the formula.\newline The object of this paper is not so much
to tell if this model is good or not or to specify the field of the validity
this model, etc... This study is simply and wholly devoted to a correction of
the Inglis-Belyaev formula in the light of its fundamental hypothesis.

\section{Hypothesis of the two-quasiparticle excitations or the cranking
Inglis-Belyaev formula.\label{section hyp}}

\subsection{Without pairing correlations}

The mass (or vibrational) parameters are given by the Inglis formula \cite{1},
\cite{2}:%

\begin{equation}
D_{ij}\left\{  \beta_{1},.,\beta_{n}\right\}  =2\hbar^{2}\sum_{M\neq0}%
\frac{\left\langle O\right\vert \partial\text{ }/\partial\beta_{i}\left\vert
M\right\rangle \left\langle M\right\vert \partial\text{ }/\partial\beta
_{j}\left\vert O\right\rangle }{E_{M}-E_{O}} \label{massparameters}%
\end{equation}
Where $\left\vert O\right\rangle ,\left\vert M\right\rangle $ are respectively
the ground state and the excited states of the nucleus. The quantities
$E_{M},E_{O}$ are the associated eigenenergies. In the independent-particle
model, whenever the state of the nucleus is assumed to be a Slater determinant
(built on single-particle states of the nucleons), the ground state$\left\vert
O\right\rangle $ will be of course the one where all the particles occupy the
lowest states. The excited states $\left\vert M\right\rangle $ will be
approached by the one particle-one hole configurations. In that case, Eq.
(\ref{massparameters}) becomes:%
\begin{equation}
D_{ij}\left\{  \beta_{1},.,\beta_{n}\right\}  =2\hbar^{2}%
{\textstyle\sum\limits_{l>\lambda,k<\lambda}}
\frac{\left\langle k\right\vert \dfrac{\partial}{\partial\beta_{i}}\left\vert
l\right\rangle \left\langle l\right\vert \dfrac{\partial}{\partial\beta_{i}%
}\left\vert k\right\rangle }{\left(  \epsilon_{l}-\epsilon_{k}\right)  }
\label{unoe}%
\end{equation}
where $\left\{  \beta_{1},.,\beta_{n}\right\}  $ or in short $\left\{
\beta\right\}  $ is a set of deformation parameters. The single particle
states $\left\vert l\right\rangle ,\left\vert k\right\rangle $ and single
particle energies $\epsilon_{l},\epsilon_{k}$ are given by the Schrodinger
equation of the independent-particle model \cite{8}, i.e. $H_{sp}\left\vert
\nu\right\rangle =\epsilon_{\nu}\left\vert \nu\right\rangle $, where $H_{sp}%
$is the single-particle Hamiltonian). At last $\lambda$ is the Fermi
level\newline\newline Using the properties $\left\langle \nu\right\vert
\partial/\partial\beta\left\vert \mu\right\rangle =\left\langle \nu\right\vert
\left[  \partial/\partial\beta,H_{sp}\right]  \left\vert \mu\right\rangle
/\left(  \epsilon_{\nu}-\epsilon_{\mu}\right)  $ and $\left[  \partial
/\partial\beta,H_{sp}\right]  =\partial H_{sp}/\partial\beta$ Eq.(\ref{unoe}) becomes%

\begin{equation}
D_{ij}\left\{  \beta_{1},.,\beta_{n}\right\}  =2\hbar^{2}%
{\textstyle\sum\limits_{l>\lambda,k<\lambda}}
\dfrac{\left\langle k\right\vert \dfrac{\partial H_{sp}}{\partial\beta_{i}%
}\left\vert l\right\rangle \left\langle l\right\vert \dfrac{\partial H_{sp}%
}{\partial\beta_{i}}\left\vert k\right\rangle }{\left(  \epsilon_{l}%
-\epsilon_{k}\right)  ^{3}} \label{massparameterssingle2}%
\end{equation}
$H_{sp}$ is the single-particle Hamiltonian and $\lambda$ is the Fermi level.

\subsection{With pairing correlations, hypothesis of the two-quasiparticle
excitations states}

It must be noted that in Eq. (\ref{massparameterssingle2}) the denominator
$\epsilon_{l}-\epsilon_{k}$ vanishes in the case where the Fermi level
coincides with two or more degenerate levels. This is the major drawback of
the formula. It is possible to overcome this difficulty by taking into account
the pairing correlations. This can be achieved through the BCS\ approximation
by the following replacements in Eq. (\ref{massparameters}):\newline\qquad i)
the ground state $\left\vert O\right\rangle $ by the BCS state $\left\vert
BCS\right\rangle .$\newline\qquad ii) the excited states $\left\vert
M\right\rangle $ by the two-quasiparticle excitations states $\left\vert
\nu,\mu\right\rangle =\alpha_{\nu}^{+}\alpha_{\mu}^{+}\left\vert
BCS\right\rangle $ (here we consider only the even-even nuclei).\newline\qquad
iii) the energy $E_{O\text{ }}$ by $E_{BCS}$ and $E_{M}$ by the energy of the
two quasiparticles, i.e., by $E_{\nu}+E_{\mu}+E_{BCS}$. The BCS state is
defined from the "true" vacuum $\left\vert 0\right\rangle $ by: $\left\vert
BCS\right\rangle =\Pi_{k}\left(  u_{k}+\upsilon_{k}a_{k}^{+}a_{\overline{k}%
}^{+}\right)  \left\vert 0\right\rangle $.\newline%
\begin{equation}
D_{ij}\left\{  \beta_{1},.,\beta_{n}\right\}  =2\hbar^{2}\sum_{\nu,\mu}%
\frac{\left\langle BCS\right\vert \partial\text{ }/\partial\beta_{i}\left\vert
\nu,\mu\right\rangle \left\langle \nu,\mu\right\vert \partial\text{ }%
/\partial\beta_{j}\left\vert BCS\right\rangle }{E_{\nu}+E_{\mu}}
\label{bcsformula}%
\end{equation}
Where $\left(  u_{\nu},\upsilon_{\mu}\right)  $ are the usual amplitudes of
probability and%
\begin{equation}
E_{\nu}=\sqrt{\left(  \epsilon_{\nu}-\lambda\right)  ^{2}+\Delta^{2}}
\label{qp}%
\end{equation}
is the so-called quasiparticle energy.\newline As shown by Belyaev \cite{6} or
as detailed in appendix tha above formula can be written in an other form:%
\begin{equation}
D_{ij}\left\{  \beta_{1},.,\beta_{n}\right\}  =2\hbar^{2}%
{\displaystyle\sum_{\nu}}
{\displaystyle\sum_{\mu\neq\nu}}
\left(  u_{\nu}\upsilon_{\mu}+u_{\mu}\upsilon_{\nu}\right)  ^{2}%
\dfrac{\left\langle \nu\right\vert \dfrac{\partial}{\partial\beta_{i}%
}\left\vert \mu\right\rangle \left\langle \mu\right\vert \dfrac{\partial
}{\partial\beta_{j}}\left\vert \nu\right\rangle }{E_{\nu}+E_{\mu}}+2\hbar
^{2}\underset{\nu}{%
{\displaystyle\sum}
}\frac{1}{2E_{\nu}}\frac{1}{\upsilon_{\nu}^{2}}\frac{\partial u_{\nu}%
}{\partial\beta_{i}}\frac{\partial u_{\nu}}{\partial\beta_{j}} \label{dudu}%
\end{equation}
Beside this formula, there is an other more convenient formulation due to Bes
\cite{7} modified slightly by the authors of Ref. \cite{1} where the
derivatives $\partial u_{\nu}/\partial\beta_{i},\partial u_{\nu}/\partial
\beta_{j}$ of Eq.(\ref{dudu}) are explitly performed (see also the details in
the appendix of the present paper):
\begin{equation}
D_{ij}\left\{  \beta_{1},.,\beta_{n}\right\}  =2\hbar^{2}%
{\displaystyle\sum_{\nu}}
{\displaystyle\sum_{\mu\neq\nu}}
\frac{\left(  u_{\nu}\upsilon_{\mu}+u_{\mu}\upsilon_{\nu}\right)  ^{2}%
}{\left(  E_{\nu}+E_{\mu}\right)  ^{3}}\left\langle \nu\right\vert
\dfrac{\partial H_{sp}}{\partial\beta_{i}}\left\vert \mu\right\rangle
\left\langle \mu\right\vert \dfrac{\partial H_{sp}}{\partial\beta_{j}%
}\left\vert \nu\right\rangle +2\hbar^{2}\underset{\nu}{%
{\displaystyle\sum}
}\dfrac{\Delta^{2}}{8E_{\nu}^{5}}R_{i}^{\nu}R_{j}^{\nu} \label{with}%
\end{equation}
here the most important quantity concerned by the subject of this paper is
$R_{i}^{\nu}$ (once again see formula (\ref{rik}) in appendix how this
quantity is obtained):%
\begin{equation}
R_{i}^{\nu}=-\left\langle \nu\right\vert \frac{\partial H_{sp}}{\partial
\beta_{i}}\left\vert \nu\right\rangle +\dfrac{\partial\lambda}{\partial
\beta_{i}}+\frac{\left(  \epsilon_{\nu}-\lambda\right)  }{\Delta}%
\dfrac{\partial\Delta}{\partial\beta_{i}} \label{mumu}%
\end{equation}
The two quantities of the r.h.s of Eq. (\ref{dudu}) and (\ref{with}) are in
the adopted order, the so-called "non-diagonal" and the "diagonal" parts of
the mass parameters. The derivatives are contained in the above diagonal term
$R_{i}^{\nu}$. In other papers, the cranking formula is usually cast under a
slightly different form. \newline\newline All these formulae (\ref{bcsformula}%
), (\ref{dudu}), (\ref{with}) and others are equivalent.\newline The
derivatives contained in Eq (\ref{mumu}) can be then actually calculated as in
the Ref. \cite{11}, \cite{1} with the help of the following formulae.
\begin{gather}
\frac{\partial\lambda}{\partial\beta_{i}}=\frac{ac_{\beta_{i}}+bd_{\beta_{i}}%
}{a^{2}+b^{2}}\label{difflamda}\\
\frac{\partial\Delta}{\partial\beta_{i}}=\frac{bc_{\beta_{i}}-ad_{\beta_{i}}%
}{a^{2}+b^{2}} \label{diffdelta}%
\end{gather}
with%
\begin{gather}
a=\underset{\nu}{%
{\displaystyle\sum}
}\Delta E_{\nu}^{-3},\text{ \ }b=\underset{\nu}{%
{\displaystyle\sum}
}(\epsilon_{\nu}-\lambda)E_{\nu}^{-3},\label{ab}\\
c_{\beta_{i}}=\underset{\nu}{%
{\displaystyle\sum}
}\Delta\left\langle \nu\right\vert \dfrac{\partial H_{sp}}{\partial\beta_{i}%
}\left\vert \nu\right\rangle E_{\nu}^{-3},\text{ \ }d_{\beta_{i}}%
=\underset{\nu}{%
{\displaystyle\sum}
}(\epsilon_{\nu}-\lambda)\left\langle \nu\right\vert \dfrac{\partial H_{sp}%
}{\partial\beta_{i}}\left\vert \nu\right\rangle E_{\nu}^{-3} \label{cgdg}%
\end{gather}
These equations can be easily derived through the well known properties of the
implicit functions. In the following the expression "the derivatives" means
simply the both derivatives given by Eq. (\ref{difflamda}) and
(\ref{diffdelta}).

In the simple BCS theory the gap parameters $\Delta$ and the Fermi level
$\lambda$ are solved from the following BCS equations (\ref{bcs1}) and
(\ref{bcs2}) as soon as the single-particle spectrum $\left\{  \epsilon_{\nu
}\right\}  $ is known.%
\begin{equation}
\dfrac{2}{G}=\underset{\nu=1}{\overset{N_{P}}{\sum}}\frac{1}{\sqrt{\left(
\epsilon_{\nu}-\lambda\right)  ^{2}+\Delta^{2}}} \label{bcs1}%
\end{equation}%
\begin{equation}
N\text{ or }Z=\underset{\nu=1}{\overset{N_{P}}{\sum}}\left(  1-\frac
{\epsilon_{\nu}-\lambda}{\sqrt{\left(  \epsilon_{\nu}-\lambda\right)
^{2}+\Delta^{2}}}\right)  \label{bcs2}%
\end{equation}
($N_{P}$ is the number of pairs of particles in numerical calculations). Of
course, from equations (\ref{bcs1}) and (\ref{bcs2}) the deformation
dependence of the eigenenergies $\epsilon_{\nu}(\beta)$ involves the ones of
$\Delta$ and $\lambda$.\newline Formally, the solution of Eq.(\ref{bcs1}) and
(\ref{bcs2}) amounts to express $\Delta$ and $\lambda$ as functions of the set
of the energy levels $\left\{  \epsilon_{\nu}\right\}  $.

\subsection{\bigskip Paradox of the formula in an umpaired
system\label{paradox}}

It is well known that the BCS equations have non-trivial solutions only above
a critical value of the strength $G$ of the pairing interaction. The trivial
solution corresponds theoretically to the value $\Delta=0$ of an unpaired
system. In this case, the mass parameters given by (\ref{dudu}) or
(\ref{with}) must reduce to the ones of the formula
(\ref{massparameterssingle2}), i.e. the cranking formula of the
independent-particle model. Indeed, when $\Delta=0$ it is quite clear that:

$E_{\nu}=\sqrt{\left(  \epsilon_{\nu}-\lambda\right)  ^{2}+\Delta^{2}%
}\rightarrow E_{\nu}=\left\vert \epsilon_{\nu}-\lambda\right\vert $

$u_{\nu},\upsilon_{\nu}\rightarrow$ $0$ $or$ $1$ therefore in Eq.(\ref{with})
$\left(  u_{\nu}\upsilon_{\mu}+u_{\mu}\upsilon_{\nu}\right)  ^{2}\rightarrow$
$0$ $or$ $1$

In accordance with the above assumption $\left(  u_{\nu}\upsilon_{\mu}+u_{\mu
}\upsilon_{\nu}\right)  ^{2}=$ $0$ $or$ $1$, we can define $\nu$ and $\mu$ in
a such way $\epsilon_{\nu}>\lambda$ and $\epsilon_{\mu}<\lambda$ therefore
$E_{\nu}+E_{\mu}=\left\vert \epsilon_{\nu}-\lambda\right\vert +\left\vert
\epsilon_{\mu}-\lambda\right\vert $ $=\epsilon_{\nu}-\lambda+\lambda
-\epsilon_{\mu}=\epsilon_{\nu}-\epsilon_{\mu}$ \newline so that it is easy to
see that the non-diagonal part of the right hand side of \ Eq.(\ref{with})
reduces effectively to Eq. (\ref{massparameterssingle2}), i.e.:

$2\hbar^{2}%
{\displaystyle\sum_{\nu}}
{\displaystyle\sum_{\mu\neq\nu}}
\frac{\left(  u_{\nu}\upsilon_{\mu}+u_{\mu}\upsilon_{\nu}\right)  ^{2}%
}{\left(  E_{\nu}+E_{\mu}\right)  ^{3}}\left\langle \nu\right\vert
\tfrac{\partial H_{sp}}{\partial\beta_{i}}\left\vert \mu\right\rangle
\left\langle \mu\right\vert \tfrac{\partial H_{sp}}{\partial\beta_{j}%
}\left\vert \nu\right\rangle \rightarrow2\hbar^{2}%
{\textstyle\sum\limits_{\nu>\lambda,\mu<\lambda}}
\tfrac{\left\langle \nu\right\vert \tfrac{\partial H_{sp}}{\partial\beta_{i}%
}\left\vert \mu\right\rangle \left\langle \mu\right\vert \tfrac{\partial
H_{sp}}{\partial\beta_{i}}\left\vert \nu\right\rangle }{\left(  \epsilon_{\nu
}-\epsilon_{\mu}\right)  ^{3}}$\newline This implies the important fact that
in this limit ($\Delta\rightarrow0$), \textit{the diagonal part (i.e. the
second term) of the r.h.s. of Eq. (\ref{with}) must vanish}, i.e. in other words:

$2\hbar^{2}\underset{\nu}{%
{\displaystyle\sum}
}\tfrac{\Delta^{2}}{8E_{\nu}^{5}}R_{i}^{\nu}R_{j}^{\nu}$ $\rightarrow0$ when
$\Delta\rightarrow0$\newline However in practice in some rare cases of the
pairing phase transtion this does not occur because it happens that this term
diverges near the breakdown of the pairing correlations, i.e., in practice for
very small values of $\Delta(\sim0)$ (see numerical example in the text
below). This constitutes really a contradiction and a paradox in this
formula.\newline In the quantity $R_{i}^{\nu}$ of Eq (\ref{mumu}) the diagonal
matrix elements $\left\langle \nu\right\vert \partial H_{sp}/\partial\beta
_{i}\left\vert \nu\right\rangle $ are finite and relatively small, it is then
clear that it is the derivatives $\partial\Delta/\partial\beta_{i}$ and
$\partial\lambda/\partial\beta_{i}$ which cause the problem. These features
have been checked in numerical calculations. In this respect, the formulae
(\ref{difflamda}) and (\ref{diffdelta}) which give these derivatives are
subject to a major drawback because their common denominator, i.e.
$a^{2}+b^{2}$ can accidentally cancel. Let us study briefly this situation. In
effect, this can be easily explained because in unpaired situation we must
have $\Delta\sim0,$ involving $a\sim0$ in Eq. (\ref{ab}). In addition, $b$ is
defined as a sum of postive and negative values depending on whether the terms
are below or above the Fermi level. Therefore,\ it could happen accidentally
that $b\sim0$ in Eq (\ref{ab}) involving serious drawbacks or at least
numerical instabilities.

\section{Quantities such as $\Delta$ and $\lambda$ are not consistent with the
hypothesis of the Inglis-Belyaev formula. \label{correction}}

\subsection{Basic hypothesis of the Inglis-Belyaev formula and simplification
of the formula}

In the independent-particle approximation the contributions to the mass
parameters are simply due to one particle-one hole excitations. Thus in the
formulae (\ref{unoe}) or (\ref{massparameterssingle2}) the particle-hole
excitations are denoted by the single-particle states $k$ and $l$. When the
pairing correlations are taken into account, the contributions are supposed
due only to two-quasiparticle excitations states $\left(  \nu,\overline{\mu
}\right)  $ $\left\{  \mu\neq\nu\right\}  $ in Eq. which gives rise to the
first term of Eq.(\ref{with}). The second term of this formula is due to the
derivatives of the probability amplitudes and has to be interpreted as two
quasiparticle excitations of the type $\left(  \nu,\overline{\nu}\right)  $.
However this is not true for all the terms entering into the product of the
quantities $R_{i}^{\nu},R_{j}^{\nu}$ . Let us re-focus onto the formula
(\ref{with}) in which we will replace in the second sum the quantity $\Delta$
by its equivalent from the identity $\Delta=2u_{\nu}\upsilon_{\nu}E_{\nu}$.
After simplification of the coefficient of $R_{i}^{\nu}R_{j}^{\nu}$ we
obtain:
\begin{equation}
D_{ij}\left\{  \beta_{1},.,\beta_{n}\right\}  =2\hbar^{2}%
{\displaystyle\sum_{\nu}}
{\displaystyle\sum_{\mu\neq\nu}}
\frac{\left(  u_{\nu}\upsilon_{\mu}+u_{\mu}\upsilon_{\nu}\right)  ^{2}%
}{\left(  E_{\nu}+E_{\mu}\right)  ^{3}}\left\langle \nu\right\vert
\dfrac{\partial H_{sp}}{\partial\beta_{i}}\left\vert \mu\right\rangle
\left\langle \mu\right\vert \dfrac{\partial H_{sp}}{\partial\beta_{j}%
}\left\vert \nu\right\rangle +2\hbar^{2}\underset{\nu}{%
{\displaystyle\sum}
}\dfrac{\left(  2u_{\nu}\upsilon_{\nu}\right)  ^{2}}{8E_{\nu}^{3}}R_{i}^{\nu
}R_{j}^{\nu} \label{mp}%
\end{equation}
The fundamental point is in this way it is clear that all the quantities in
Eq. (\ref{mp}) are associated to quasiparticle states $\nu$ and $\mu$ except
the derivative of $\Delta$ and $\lambda.$ \ Quantities such as $\Delta$ and
$\lambda$ appearing in $\left(  R_{i}^{\nu},R_{j}^{\nu}\right)  $ (see Eq.
(\ref{mumu})) which are deduced from Eq. (\ref{bcs1})-(\ref{bcs2}) are due to
all the spectrum, they are clearly not specifically linked to these two
particular states (otherwise indices $\nu$ and $\mu$ should appear with these
quantities). Therefore they cannot be really considered as contributions due
to two quasiparticle excitation states which is the basic hypothesis of the
Inglis-Belyaev formula. Therefore they cannot be taken into account.\newline
With this additional assumption, the element $R_{i}^{\nu}$ must reduce to
nothing but a simple matrix element:%

\begin{equation}
R_{i}^{\nu}=-\left\langle \nu\right\vert \frac{\partial H_{sp}}{\partial
\beta_{i}}\left\vert \nu\right\rangle
\end{equation}
Consequently this contributes to simplify greatly the formula (\ref{mp}) which
becomes:%
\begin{equation}
D_{ij}\left\{  \beta_{1},.,\beta_{n}\right\}  =2\hbar^{2}%
{\displaystyle\sum_{\nu}}
{\displaystyle\sum_{\mu\neq\nu}}
\left(  u_{\nu}\upsilon_{\mu}+u_{\mu}\upsilon_{\nu}\right)  ^{2}%
\frac{\left\langle \nu\right\vert \dfrac{\partial H_{sp}}{\partial\beta_{i}%
}\left\vert \mu\right\rangle \left\langle \mu\right\vert \dfrac{\partial
H_{sp}}{\partial\beta_{j}}\left\vert \nu\right\rangle }{\left(  E_{\nu}%
+E_{\mu}\right)  ^{3}}+2\hbar^{2}\underset{\nu}{%
{\displaystyle\sum}
}\left(  2u_{\nu}\upsilon_{\nu}\right)  ^{2}\frac{\left\langle \nu\right\vert
\dfrac{\partial H_{sp}}{\partial\beta_{i}}\left\vert \nu\right\rangle ^{2}%
}{\left(  2E_{\nu}\right)  ^{3}} \label{mpp}%
\end{equation}
It is to be noted that the missing term $(\mu=\nu)$ in the double sum is
precisely the contribution of the simple sum of the r.h.s of Eq. (\ref{mpp}).
Therefore, the formula (\ref{mpp}) can be reformulated in a compact form:%

\begin{equation}
D_{ij}\left\{  \beta_{1},.,\beta_{n}\right\}  =2\hbar^{2}%
{\displaystyle\sum_{\nu}}
{\displaystyle\sum_{\mu}}
\left(  u_{\nu}\upsilon_{\mu}+u_{\mu}\upsilon_{\nu}\right)  ^{2}%
\frac{\left\langle \nu\right\vert \dfrac{\partial H_{sp}}{\partial\beta_{i}%
}\left\vert \mu\right\rangle \left\langle \mu\right\vert \dfrac{\partial
H_{sp}}{\partial\beta_{j}}\left\vert \nu\right\rangle }{\left(  E_{\nu}%
+E_{\mu}\right)  ^{3}}%
\end{equation}
In this more "symmetric" form, this formula looks like more naturally to the
Inglis-Belyaev formula of the moments of inertia which does not contain
dependence on the derivatives of $\Delta$ and $\lambda$.

The "new" formula corrects the previous paradox because in the case of the
phase transition $\Delta\rightarrow0$, we will have in this limit for the
diagonal terms $\nu=\mu,$ $u_{\nu}\upsilon_{\nu}+u_{\nu}\upsilon_{\nu}=0$ for
any $\nu$ and due to the fact that the corresponding matrix element
$\left\langle \nu\right\vert \partial H_{sp}/\partial\beta_{i}\left\vert
\nu\right\rangle $ is finite the second term of Eq. (\ref{mpp}) tends
uniformly toward zero so that Eq. (\ref{mpp}) reduces in this case to the
equation of the unpaired system (\ref{massparameterssingle2}) without any problem.

\section{Illustration of the application of the Inglis Belyaev formula in the
case where the singularity occurs}

\begin{figure}[ptbh]
\includegraphics[angle=0,width=140mm,keepaspectratio]{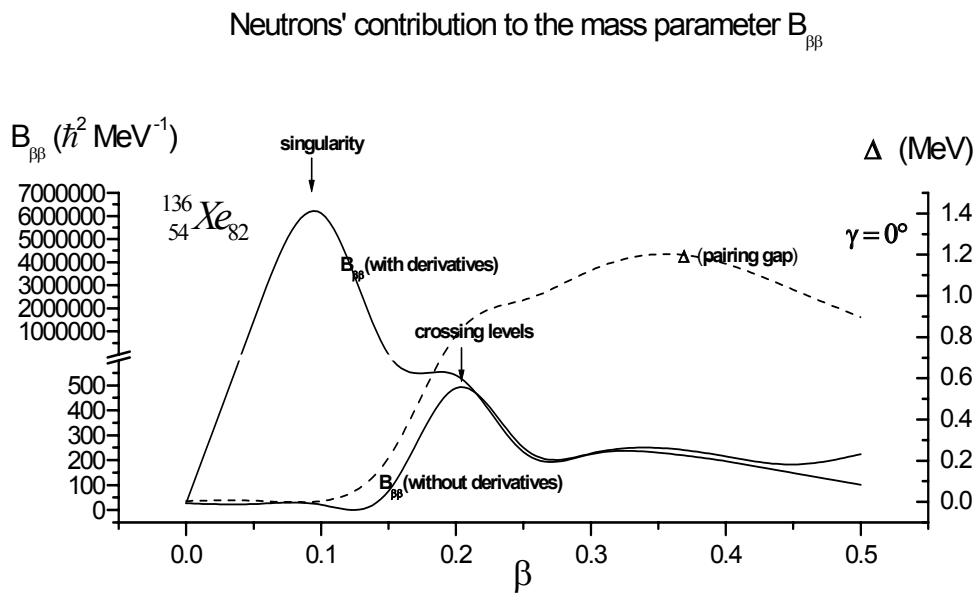}\caption{Neutron
contribution to the mass parameters $B_{\beta\beta}$ for the magic nucleus
$_{54}^{136}Xe_{82}$; The calculations are performed within the cranking
formula including the derivatives and for the same formula without derivating;
Note the quasi divergence (singularity) of the version with derivatives near
the deformation $\beta=0.1$}%
\label{fig1}%
\end{figure}

This is illustrated in Fig. \ref{fig1} by the behaviour of the vibrational
parameter $B_{\beta\beta}(\beta,\gamma=0)$ as a function of the Bohr parameter
in the case of the magic nuclei $_{54}^{136}$Xe$_{82}$. These calculations
have been performed for the both formulae (\ref{with}) and (\ref{mpp}), i.e.,
respectively with and without the derivatives $\partial\lambda/\partial\beta$
and $\partial\Delta/\partial\beta$. The resonance (singularity) $B_{\beta
\beta}\sim7000000\hbar^{2}MeV^{-1}$ occurs near the deformation $\beta=0.09$
for the formula with the derivatives. This happens even if $\Delta$ is very
close to $0.$ Between $\beta=0$ and $\beta=0.15$ the formula without
derivatives gives small (finite) values ($B_{\beta\beta}\sim25\hbar
^{2}MeV^{-1}$). These very small values of the independent particle model are
due to the collapse of the pairing correlations. In addition, during the phase
transition, i.e., for $0.1\lesssim\Delta\lesssim0.2,$ the vibrational
parameters increase up to the important value $B_{\beta\beta}\sim500\hbar
^{2}MeV^{-1}$. We have checked that this is due to a pseudo crossing levels
near the Fermi level. However, in this respect we have futhermore checked
carefully that there is absolutely no crossing levels near the singularity.
Thus the singularity is not a consequence of a crossing levels as it is often
claimed \cite{10}. As said before the explanation comes from the fact that in
Eq. (\ref{difflamda}) and (\ref{diffdelta}) the denominator simply cancels.
This demonstrates the weakness of the old formula (\ref{with}) with respect to
that proposed in this paper, that is Eq. (\ref{mpp}).\newline

\section{Conclusion}

In some rare but important illustrative cases the application of the
Inglis-Belyaev formula to the mass parameters reveals incontestable weaknesses
in the limit of unpaired systems $\Delta\rightarrow0$. In effect, this formula
leads straightforwardly to a major contradiction, that is, not only it does
not reduce to the one of the unpaired system in the case $\Delta=0$ (which is
already a contradictory fact) but even gives unphysical (singular) values. It
has been reported in the litterature that self-consistent calculations meet
also the same kind of problems (see text). After extensive calculations within
the Inglis-Belyaev formula, we realized that these problems are inherent to a
spurious presence of the derivatives of $\Delta$ and $\lambda$ in the formula.
This led us to "revise" the conception of this formula simply by removing the
derivatives which are not consistent with the basic hypothesis of the formula,
that is to say with two quasiparticle excitation states. This is the reason
why our proposal cannot be considered as a simple recipe to the limit
$\Delta=0$ but as a well founded rectification of the formula which is thus no
more subject to the cited problems and reduces naturally to that of the
unpaired system in the limit $\Delta\rightarrow0$.

\appendix

\section{The cranking formula with pairing correlations}

We have to calculate the matrix element of the type $\left\langle
n,m\right\vert \partial$ $/\partial\beta_{i}\left\vert BCS\right\rangle $
which appears in Eq. (\ref{bcsformula}) of the text, i.e.:

$D_{ij}\left\{  \beta_{1},.,\beta_{n}\right\}  =2\hbar^{2}\sum_{\nu,\mu}%
\frac{\left\langle BCS\right\vert \partial\text{ }/\partial\beta_{i}\left\vert
\nu,\mu\right\rangle \left\langle \nu,\mu\right\vert \partial\text{ }%
/\partial\beta_{j}\left\vert BCS\right\rangle }{E_{\nu}+E_{\mu}}$\newline
keeping in mind however that the differential operator acts not only on the
wave functions of the BCS state but also on the occupations probabilities
$u_{k},\upsilon_{k}$ (of the BCS state) which also depend on the deformation
parameter $\beta_{i}$ we have to write.

$\dfrac{\partial}{\partial\beta_{i}}=\left(  \dfrac{\partial}{\partial
\beta_{i}}\right)  _{wave\text{ }func}+\left(  \dfrac{\partial}{\partial
\beta_{i}}\right)  _{occup.prob}$\newline We must therefore to evaluate
successively two types of matrix elements

\subsection{Calculation of the first type of matrix elements\label{first type}%
}

For one particle operator we have in second quantization
representation:\newline$\left(  \dfrac{\partial}{\partial\beta_{i}}\right)
_{wave\text{ }func}=\sum_{\nu,\mu}\left\langle \nu\right\vert \dfrac{\partial
}{\partial\beta_{i}}\left\vert \mu\right\rangle a_{\nu}^{+}a_{\mu}^{{}}%
$\newline Applying this operator on the paired system and using the inverse of
the Bogoliubov-Valatin transformation:\newline$a_{\nu}=(u_{\nu}\alpha_{\nu
}+\upsilon_{\nu}\alpha_{\overline{\nu}}^{+}),a_{\nu}^{+}=(u_{\nu}\alpha_{\nu
}^{+}+\upsilon_{\nu}\alpha_{\overline{\nu}})$\newline We find:%
\begin{equation}
\left(  \dfrac{\partial}{\partial\beta_{i}}\right)  _{wave\text{ }%
func}\left\vert BCS\right\rangle =\sum_{\nu,\mu}\left\langle \nu\right\vert
\dfrac{\partial}{\partial\beta_{i}}\left\vert \mu\right\rangle a_{\nu}%
^{+}a_{\mu}^{{}}\left\vert BCS\right\rangle \label{twoqp}%
\end{equation}
$=\sum_{\nu,\mu}\left\langle \nu\right\vert \dfrac{\partial}{\partial\beta
_{i}}\left\vert \mu\right\rangle (u_{\nu}\alpha_{\nu}^{+}+\upsilon_{\nu}%
\alpha_{\overline{\nu}})(u_{\mu}\alpha_{\mu}+\upsilon_{\mu}\alpha
_{\overline{\mu}}^{+})\left\vert BCS\right\rangle =\sum_{\nu,\mu}\left\langle
\nu\right\vert \dfrac{\partial}{\partial\beta_{i}}\left\vert \mu\right\rangle
(u_{\nu}\alpha_{\nu}^{+}+\upsilon_{\nu}\alpha_{\overline{\nu}})\upsilon_{\mu
}\alpha_{\overline{\mu}}^{+}\left\vert BCS\right\rangle $ because $\alpha
_{\mu}\left\vert BCS\right\rangle =0$\newline Therefore%

\begin{equation}
\left(  \dfrac{\partial}{\partial\beta_{i}}\right)  _{wave\text{ }%
func}\left\vert BCS\right\rangle =\sum_{\nu,\mu}\left\langle \nu\right\vert
\dfrac{\partial}{\partial\beta_{i}}\left\vert \mu\right\rangle \left\{
u_{\nu}\alpha_{\nu}^{+}\upsilon_{\mu}\alpha_{\overline{\mu}}^{+}\left\vert
BCS\right\rangle +\upsilon_{\nu}\alpha_{\overline{\nu}}\upsilon_{\mu}%
\alpha_{\overline{\mu}}^{+}\left\vert BCS\right\rangle \right\}  \label{vnsh}%
\end{equation}
\newline We must notice that for the term $\nu=\mu$ we will have the
contribution $\left\langle \nu\right\vert \dfrac{\partial}{\partial\beta_{i}%
}\left\vert \nu\right\rangle \left\{  u_{\nu}\upsilon_{\nu}\alpha_{\nu}%
^{+}\alpha_{\overline{\nu}}^{+}\left\vert BCS\right\rangle +\upsilon_{\nu}%
^{2}\left\vert BCS\right\rangle \right\}  $ which is a mixing of a two
quasparticle-state with a BCS state. Because the state given by Eq.
(\ref{twoqp}) must represent only two quasiparticle excitation, we have to
exclude the contribution due to the term $\nu=\mu$ from the sum of this
equation. This restriction leads to the following formula:%

\begin{equation}
\left(  \dfrac{\partial}{\partial\beta_{i}}\right)  _{wave\text{ }%
func}\left\vert BCS\right\rangle =\sum_{\nu\neq\mu}\left\langle \nu\right\vert
\dfrac{\partial}{\partial\beta_{i}}\left\vert \mu\right\rangle (u_{\nu
}\upsilon_{\mu}\alpha_{\nu}^{+}\alpha_{\overline{\mu}}^{+})\left\vert
BCS\right\rangle \label{op}%
\end{equation}
It will be noted that the term $\upsilon_{\nu}\alpha_{\overline{\nu}}%
\upsilon_{\mu}\alpha_{\overline{\mu}}^{+}\left\vert BCS\right\rangle $
vanishes for $\nu\neq\mu$ in the r.h.s of Eq. (\ref{vnsh}). We then calculate
then the first type of matrix elements:\newline%
\begin{equation}
I_{1}=\left\langle n,m\right\vert \sum_{\nu\neq\mu}\left\langle \nu\right\vert
\dfrac{\partial}{\partial\beta_{i}}\left\vert \mu\right\rangle u_{\nu}%
\upsilon_{\mu}\alpha_{\nu}^{+}\alpha_{\overline{\mu}}^{+}\left\vert
BCS\right\rangle \label{firsttype}%
\end{equation}
\newline The above form of the formula suggests that the excited states must
be of the form $\left\vert n,m\right\rangle =\alpha_{k}^{+}\alpha
_{\overline{l}}^{+}\left\vert BCS\right\rangle =\left\vert k,\overline
{l}\right\rangle $.\newline We obtain then:\newline$I_{1}=\left\langle
BCS\right\vert \alpha_{\overline{l}}^{{}}\alpha_{k}^{{}}\sum_{\nu\neq\mu
}\left\langle \nu\right\vert \dfrac{\partial}{\partial\beta_{i}}\left\vert
\mu\right\rangle u_{\nu}\upsilon_{\mu}\alpha_{\nu}^{+}\alpha_{\overline{\mu}%
}^{+}\left\vert BCS\right\rangle $\newline$=\sum_{\nu\neq\mu}\left\langle
\nu\right\vert \dfrac{\partial}{\partial\beta_{i}}\left\vert \mu\right\rangle
u_{\nu}\upsilon_{\mu}\left\langle BCS\right\vert \alpha_{\overline{l}}^{{}%
}\alpha_{k}^{{}}\alpha_{\nu}^{+}\alpha_{\overline{\mu}}^{+}\left\vert
BCS\right\rangle $\newline We use the following usual fermions anticommutation
relations:\newline$\left\{  \alpha_{k}^{{}},\alpha_{l}^{{}}\right\}  =\left\{
\alpha_{k}^{+},\alpha_{l}^{+}\right\}  =0,$ \ \ \ $\left\{  \alpha_{k}^{{}%
},\alpha_{l}^{+}\right\}  =\delta_{kl}$\newline Thus the quantity between
brakets of the BCS sate gives:\newline$\left\langle BCS\right\vert
\alpha_{\overline{l}}^{{}}\alpha_{k}^{{}}\alpha_{\nu}^{+}\alpha_{\overline
{\mu}}^{+}\left\vert BCS\right\rangle =\left(  \delta_{l\mu}\delta_{\nu
k}-\delta_{\overline{\mu}k}\delta_{\nu\overline{l}}\right)  $\newline We
obtain:\newline$I_{1}=\sum_{\nu\neq\mu}\left\langle \nu\right\vert
\dfrac{\partial}{\partial\beta_{i}}\left\vert \mu\right\rangle u_{\nu}%
\upsilon_{\mu}\left(  \delta_{l\mu}\delta_{\nu k}-\delta_{\overline{\mu}%
k}\delta_{\nu\overline{l}}\right)  =\left\langle k\right\vert \dfrac{\partial
}{\partial\beta_{i}}\left\vert l\right\rangle u_{k}\upsilon_{l}-\left\langle
\overline{l}\right\vert \dfrac{\partial}{\partial\beta_{i}}\left\vert
\overline{k}\right\rangle u_{\overline{l}}\upsilon_{\overline{k}}$ \ \ with
\ \ $k\neq l$ Because indexes of brakets must be different in
Eq(\ref{firsttype}).\newline Noting that if $\widehat{T}$ is the time-reversal
conjugation operator we must have for any operator $\hat{O}$\newline%
$\left\langle p\right\vert \hat{O}\left\vert q\right\rangle =\left\langle
\widehat{T}p\right\vert \widehat{T}\hat{O}\widehat{T}^{-1}\left\vert
\widehat{T}q\right\rangle ^{\ast}$\newline Applying this result for our case
and assuming that $\partial/\partial\beta_{i}$ is time-even, i.e. $\widehat
{T}\left(  \partial/\partial\beta_{i}\right)  \widehat{T}^{-1}=\partial
/\partial\beta_{i}$, we get:\newline$\left\langle \overline{l}\right\vert
\dfrac{\partial}{\partial\beta_{i}}\left\vert \overline{k}\right\rangle
=\left\langle \widehat{T}\overline{l}\right\vert \widehat{T}\dfrac{\partial
}{\partial\beta_{i}}\widehat{T}^{-1}\left\vert \widehat{T}\overline
{k}\right\rangle ^{\ast}=\left\langle k\right\vert \dfrac{\partial}%
{\partial\beta_{i}}\left\vert l\right\rangle $\newline Moreover, using the
usual phase convention\newline$u_{\overline{l}}=u_{l}$, $\upsilon
_{\overline{k}}=-\upsilon_{k}$\newline we deduce :\newline$I_{1}=\left\langle
k\right\vert \dfrac{\partial}{\partial\beta_{i}}\left\vert l\right\rangle
u_{k}\upsilon_{l}+\left\langle k\right\vert \dfrac{\partial}{\partial\beta
_{i}}\left\vert l\right\rangle u_{l}\upsilon_{k}=\left(  u_{k}\upsilon
_{l}+u_{l}\upsilon_{k}\right)  \left\langle k\right\vert \dfrac{\partial
}{\partial\beta_{i}}\left\vert l\right\rangle $\ \ \newline Taking into
account that the brakets states in Eq (\ref{op}) must be different, the final
result for $I_{1}$ given (\ref{firsttype}) will take the following form:%

\begin{equation}
I_{1}=\left\langle k,\overline{l}\right\vert \left(  \frac{\partial}%
{\partial\beta_{i}}\right)  _{wave\text{ }func}\left\vert BCS\right\rangle
=\left(  u_{k}\upsilon_{l}+u_{l}\upsilon_{k}\right)  \left\langle k\right\vert
\frac{\partial}{\partial\beta_{i}}\left\vert l\right\rangle \ \ with\ \ k\neq
l \label{a3}%
\end{equation}
Let be $H_{sp}$ the single-particle Hamiltonian and%

\[
H^{\prime}=%
{\textstyle\sum_{\nu,\mu}}
\left\langle \nu\right\vert \left(  H_{sp}-\lambda\right)  \left\vert
\mu\right\rangle a_{\nu}^{+}a_{\mu}-G%
{\textstyle\sum_{\nu,\mu>0}}
a_{\nu}^{+}a_{\overline{\nu}}^{+}a_{\overline{\mu}}a_{\mu}%
\]
the nuclear paired BCS Hamiltonian with the constraint on the particle number.
Writing this Hamiltonian in the well-known quasiparticles representation
$H^{\prime}=E_{BCS}+%
{\textstyle\sum_{\nu}}
E_{\nu}\alpha_{\nu}^{+}\alpha_{\nu}+residual$ $qp$ $interaction$, neglecting
(as usual) the latter term and using Eq. (\ref{op}) it is quite easy to
establish the following identity\newline%
\[
\left\langle k,\overline{l}\right\vert \left[  H^{\prime},\left(
\dfrac{\partial}{\partial\beta_{i}}\right)  _{wave\text{ }func}\right]
\left\vert BCS\right\rangle =-\left\langle k,\overline{l}\right\vert \left(
\dfrac{\partial H^{\prime}}{\partial\beta_{i}}\right)  _{wave\text{ }%
func}\left\vert BCS\right\rangle =\left(  E_{k,\overline{l}}-E_{BCS}\right)
\left\langle k,\overline{l}\right\vert \left(  \dfrac{\partial}{\partial
\beta_{i}}\right)  _{wave\text{ }func}\left\vert BCS\right\rangle
\]
where the eigenenergies $E_{k,\overline{l}}$ corresponding to the excited
states $\left\vert k,\overline{l}\right\rangle $ are given by $E_{k,\overline
{l}}=E_{BCS}+E_{k}+E_{l}$\newline so that:\newline%
\[
\left\langle k,\overline{l}\right\vert \left(  \dfrac{\partial}{\partial
\beta_{i}}\right)  _{wave\text{ }func}\left\vert BCS\right\rangle
=-\dfrac{\left\langle k,\overline{l}\right\vert \left(  \dfrac{\partial
H^{\prime}}{\partial\beta_{i}}\right)  _{wave\text{ }func}\left\vert
BCS\right\rangle }{E_{k,\overline{l}}-E_{BCS}}=-\dfrac{\left\langle
k,\overline{l}\right\vert \left(  \dfrac{\partial H^{\prime}}{\partial
\beta_{i}}\right)  _{wave\text{ }func}\left\vert BCS\right\rangle }%
{E_{k}+E_{l}}%
\]
\newline Due to the fact that the pairing strength $G$ does not depend on the
nuclear deformation, it is clear from the expression of $H^{\prime}$(in the
particles representation) that $\partial H^{\prime}/\partial\beta_{i}%
=\partial\left(  H_{sp}-\lambda\right)  /\partial\beta_{i}$ Therefore
$\left\langle k,\overline{l}\right\vert \left(  \partial H^{\prime}%
/\partial\beta_{i}\right)  _{wave\text{ }func}\left\vert BCS\right\rangle
=\left\langle k,\overline{l}\right\vert \left(  \partial H_{sp}/\partial
\beta_{i}\right)  \left\vert BCS\right\rangle -\partial\lambda/\partial
\beta_{i}\langle k,\overline{l}\left\vert BCS\right\rangle =\left\langle
k,\overline{l}\right\vert \left(  \partial H_{sp}/\partial\beta_{i}\right)
\left\vert BCS\right\rangle $\newline Here we have $\langle k,\overline
{l}\left\vert BCS\right\rangle =0$ because excited states and bcs state are
supposed orthogonal.\newline Again using the second quantization formalism
$\left(  \partial H_{sp}/\partial\beta_{i}\right)  =\sum_{\nu\neq\mu
}\left\langle \nu\right\vert \left(  \partial H_{sp}/\partial\beta_{i}\right)
\left\vert \mu\right\rangle a_{\nu}^{+}a_{\mu}^{{}}$ and performing then
exactly the same transformations as before for $\sum_{\nu\neq\mu}\left\langle
\nu\right\vert \partial/\partial\beta_{i}\left\vert \mu\right\rangle a_{\nu
}^{+}a_{\mu}$ but this time with $\sum_{\nu\neq\mu}\left\langle \nu\right\vert
\partial H_{sp}/\partial\beta_{i}\left\vert \mu\right\rangle a_{\nu}^{+}%
a_{\mu}$ we will obtain in the same manner a new form for Eq. (\ref{a3}):%

\begin{equation}
I_{1}=I_{1}(k,l)=\left\langle k,\overline{l}\right\vert \left(  \frac
{\partial}{\partial\beta_{i}}\right)  _{wave\text{ }func}\left\vert
BCS\right\rangle =-\frac{\left(  u_{k}\upsilon_{l}+u_{l}\upsilon_{k}\right)
}{E_{k}+E_{l}}\left\langle k\right\vert \frac{\partial H_{sp}}{\partial
\beta_{i}}\left\vert l\right\rangle \ \ with\ \ k\neq l \label{aa}%
\end{equation}

\subsection{Calculation of the second type of matrix
elements\label{second type}}

Recalling that the BCS state is given by: $\left\vert BCS\right\rangle
=\Pi_{k}\left(  u_{k}+\upsilon_{k}a_{k}^{+}a_{\overline{k}}^{+}\right)
\left\vert 0\right\rangle $ and differentiating this state with respect to the
probabilty amplitudes, we obtain:\newline$\left(  \dfrac{\partial}%
{\partial\beta_{i}}\right)  _{occup.prob}\left\vert BCS\right\rangle
=\sum_{\tau}(\dfrac{\partial u_{\tau}}{\partial\beta_{i}}+\dfrac
{\partial\upsilon_{\tau}}{\partial\beta_{i}}a_{\tau}^{+}a_{\overline{\tau}%
}^{+})\prod_{k\neq\tau}(u_{k}+\upsilon_{k}a_{k}^{+}a_{\overline{k}}%
^{+})\left\vert 0\right\rangle $\newline We use the evident property:\newline%
$\prod_{k\neq\tau}(u_{k}+\upsilon_{k}a_{k}^{+}a_{\overline{k}}^{+})\left\vert
0\right\rangle $ $=(u_{\tau}+\upsilon_{\tau}a_{\tau}^{+}a_{\overline{\tau}%
}^{+})^{-1}\left\vert BCS\right\rangle $\newline Therefore:\newline$\left(
\dfrac{\partial}{\partial\beta_{i}}\right)  _{occup.prob}\left\vert
BCS\right\rangle =\sum_{\tau}\left[  (\dfrac{\partial u_{\tau}}{\partial
\beta_{i}}+\dfrac{\partial\upsilon_{\tau}}{\partial\beta_{i}}a_{\tau}%
^{+}a_{\overline{\tau}}^{+})(u_{\tau}+\upsilon_{\tau}a_{\tau}^{+}%
a_{\overline{\tau}}^{+})^{-1}\right]  \left\vert BCS\right\rangle $\newline
Making an expansion of the inverse operator in $a_{\tau}^{+}a_{\overline{\tau
}}^{+}$ :\newline$\left(  \dfrac{\partial}{\partial\beta_{i}}\right)
_{occup.prob}\left\vert BCS\right\rangle =\sum_{\tau}\left[  (\dfrac{\partial
u_{\tau}}{\partial\beta_{i}}+\dfrac{\partial\upsilon_{\tau}}{\partial\beta
_{i}}a_{\tau}^{+}a_{\overline{\tau}}^{+})u_{\tau}^{-1}(1-\upsilon_{\tau
}u_{\tau}^{-1}a_{\tau}^{+}a_{\overline{\tau}}^{+}+\left(  \upsilon_{\tau
}u_{\tau}^{-1}a_{\tau}^{+}a_{\overline{\tau}}^{+}\right)  ^{2}+...)\right]
\left\vert BCS\right\rangle $\newline using the inverse of the
Bogoliubov-Valatin transformation:\newline$a_{\tau}^{+}=(u_{\tau}\alpha_{\tau
}^{+}+\upsilon_{\tau}\alpha_{\overline{\tau}})$\newline We find for the
quantity $a_{\tau}^{+}a_{\overline{\tau}}^{+}$\newline$a_{\tau}^{+}%
a_{\overline{\tau}}^{+}=(u_{\tau}\alpha_{\tau}^{+}+\upsilon_{\tau}%
\alpha_{\overline{\tau}})(u_{\overline{\tau}}\alpha_{\overline{\tau}}%
^{+}+\upsilon_{\overline{\tau}}\alpha_{\tau})$\newline$=u_{\tau}%
u_{\overline{\tau}}\alpha_{\tau}^{+}\alpha_{\overline{\tau}}^{+}+u_{\tau
}\upsilon_{\overline{\tau}}\alpha_{\tau}^{+}\alpha_{\tau}+\upsilon_{\tau
}u_{\overline{\tau}}\alpha_{\overline{\tau}}\alpha_{\overline{\tau}}%
^{+}+\upsilon_{\tau}\upsilon_{\overline{\tau}}\alpha_{\overline{\tau}}%
\alpha_{\tau}$\newline replacing in the above expression and retaining only
two creation of quasiparticles with at most products of two amplitude
probability:\newline$\left(  \dfrac{\partial}{\partial\beta_{i}}\right)
_{occup.prob}\left\vert BCS\right\rangle =\sum_{\tau}\left[  \dfrac{\partial
u_{\tau}}{\partial\beta_{i}}u_{\tau}^{-1}\left(  -\upsilon_{\tau}u_{\tau}%
^{-1}u_{\tau}u_{\overline{\tau}}\alpha_{\tau}^{+}\alpha_{\overline{\tau}}%
^{+}\right)  +u_{\tau}^{-1}\dfrac{\partial\upsilon_{\tau}}{\partial\beta_{i}%
}u_{\tau}u_{\overline{\tau}}\alpha_{\tau}^{+}\alpha_{\overline{\tau}}%
^{+}\right]  \left\vert BCS\right\rangle $\newline Noting that: $u_{\overline
{\tau}}=u_{\tau}$, $\upsilon_{\overline{\tau}}=-\upsilon_{\tau}$, we
find\newline$\left(  \dfrac{\partial}{\partial\beta_{i}}\right)
_{occup.prob}\left\vert BCS\right\rangle =\sum_{\tau}\left[  u_{\tau}%
\dfrac{\partial\upsilon_{\tau}}{\partial\beta_{i}}\newline-\upsilon_{\tau
}\dfrac{\partial u_{\tau}}{\partial\beta_{i}}\right]  \alpha_{\tau}^{+}%
\alpha_{\overline{\tau}}^{+}\left\vert BCS\right\rangle $\newline The excited
states will be necessarily here, of the following form:\newline$\left\vert
M\right\rangle =\alpha_{m}^{+}\alpha_{\overline{m}}^{+}\left\vert
BCS\right\rangle =\left\vert m,\overline{m}\right\rangle $\newline We have
therefore to calculate:\newline$I_{2}=\left\langle BCS\right\vert
\alpha_{\overline{m}}^{{}}\alpha_{m}^{{}}(u_{m}\dfrac{\partial\upsilon_{m}%
}{\partial\beta_{i}}-\upsilon_{m}\dfrac{\partial u_{m}}{\partial\beta_{i}%
})\alpha_{m}^{+}\alpha_{\overline{m}}^{+}\left\vert BCS\right\rangle $\newline
due to the normalisation of the excited states, we obtains:\newline%
$I_{2}=u_{m}\dfrac{\partial\upsilon_{m}}{\partial\beta_{i}}-\upsilon_{m}%
\dfrac{\partial u_{m}}{\partial\beta_{i}}$\newline knowing that the
normalization condition of the probability amplitudes is:\newline$u_{m}%
^{2}+\upsilon_{m}^{2}=1$\newline we find by differentiation\newline%
$2u_{m}\dfrac{\partial u_{m}}{\partial\beta_{i}}+2\upsilon_{m}\dfrac
{\partial\upsilon_{m}}{\partial\beta_{i}}=0$\newline combining these two
relations, we obtain in $I_{2}$:\newline$I_{2}=-\dfrac{1}{\upsilon_{m}}%
\dfrac{\partial u_{m}}{\partial\beta_{i}}$\newline then, the second term
reads:\newline$I_{2}=\left\langle m,\overline{m}\right\vert \left(
\frac{\partial}{\partial\beta_{i}}\right)  _{occup.prob}\left\vert
BCS\right\rangle =-\dfrac{1}{\upsilon_{m}}\dfrac{\partial u_{m}}{\partial
\beta_{i}}$\newline which can be cast as follows:%

\begin{equation}
I_{2}=I_{2}(k,l)=\left\langle k,\overline{l}\right\vert \left(  \frac
{\partial}{\partial\beta_{i}}\right)  _{occup.prob}\left\vert BCS\right\rangle
=-\frac{1}{\upsilon_{k}}\frac{\partial u_{k}}{\partial\beta_{i}}%
\ \ with\ \ k=l \label{a4}%
\end{equation}
The two matrix elements $I_{1}$ given by Eq. (\ref{aa}) and $I_{2}$ given by
Eq. (\ref{a4}). They correspond respectively to the non-diagonal $k\neq l$ and
diagonal part $k=l$ of the total contribution. Reassembling the two parts
$I_{1}$ and $I_{2}$ in only one formula, we get:\newline$I_{1}+I_{2}%
=\left\langle k,\overline{l}\right\vert \dfrac{\partial}{\partial\beta_{i}%
}\left\vert BCS\right\rangle =-\dfrac{\left(  u_{k}\upsilon_{l}+u_{l}%
\upsilon_{k}\right)  }{E_{k}+E_{l}}\left\langle k\right\vert \dfrac{\partial
H_{sp}}{\partial\beta_{i}}\left\vert l\right\rangle \left(  1-\delta
_{k,l}\right)  -\dfrac{1}{\upsilon_{k}}\dfrac{\partial u_{k}}{\partial
\beta_{i}}\delta_{kl}$\newline Replacing this quantity in Eq.
(\ref{bcsformula}) of section \ref{section hyp}, noticing that the crossed
terms $(I_{1}I_{2}$ and $I_{2}I_{1})$ cancel in the product we find:%

\begin{equation}
D_{ij}\left\{  \beta_{1},.,\beta_{n}\right\}  =2\hbar^{2}\sum_{k,l}%
\frac{\left(  u_{k}\upsilon_{l}+u_{l}\upsilon_{k}\right)  ^{2}}{\left(
E_{k}+E_{l}\right)  ^{3}}\left\langle l\right\vert \frac{\partial H_{sp}%
}{\partial\beta_{i}}\left\vert k\right\rangle \left\langle k\right\vert
\frac{\partial H_{sp}}{\partial\beta_{j}}\left\vert l\right\rangle \left(
1-\delta_{k,l}\right)  +2\hbar^{2}\sum_{k}\frac{1}{2E_{k}}\frac{1}%
{\upsilon_{k}^{2}}\frac{\partial u_{k}}{\partial\beta_{i}}\frac{\partial
u_{k}}{\partial\beta_{j}} \label{a5}%
\end{equation}
The expression\newline%
\begin{equation}
\sum_{k}\frac{1}{2E_{k}}\frac{1}{\upsilon_{k}^{2}}\frac{\partial u_{k}%
}{\partial\beta_{i}}\frac{\partial u_{k}}{\partial\beta_{j}} \label{prod}%
\end{equation}
meet in the second part of the r.h.s of the above formula (\ref{a5}) can be
further clarified. Recalling that the probabilty amplitudes are:\newline%
$u_{k}^{{}}=\left(  1/\sqrt{2}\right)  \left(  1+\varepsilon_{k}%
/\sqrt{\varepsilon_{k}^{2}+\Delta^{2}}\right)  ^{1/2}$ and $\upsilon_{k}^{{}%
}=\left(  1/\sqrt{2}\right)  \left(  1-\varepsilon_{k}/\sqrt{\varepsilon
_{k}^{2}+\Delta^{2}}\right)  ^{1/2}$\newline where:$\varepsilon_{k}%
=\epsilon_{k}-\lambda$ is the single-particle energy with respect to the Fermi
level $\lambda$, $\epsilon_{k}$ being the single particle energy. Since the
deformation dependence in $u_{k}$ appears through $\epsilon_{k},\Delta,$ and
$\lambda$, a simple differentiation of $u_{k}$ with respect to $\beta_{i}$
leads to:\newline$\dfrac{\partial u_{k}^{{}}}{\partial\beta_{i}}=\dfrac
{1}{2\sqrt{2}}\left(  1+\dfrac{\varepsilon_{k}}{\sqrt{\varepsilon_{k}%
^{2}+\Delta^{2}}}\right)  ^{-1/2}\left[  \dfrac{\partial\varepsilon_{k}%
}{\partial\beta_{i}}\left(  \varepsilon_{k}^{2}+\Delta^{2}\right)
^{-1/2}-\varepsilon_{k}\left(  \varepsilon_{k}^{2}+\Delta^{2}\right)
^{-3/2}\left(  \varepsilon_{k}\dfrac{\partial\varepsilon_{k}}{\partial
\beta_{i}}+\Delta\dfrac{\partial\Delta}{\partial\beta_{i}}\right)  \right]
$\newline multiplying by $\dfrac{1}{\upsilon_{k}}$and simplifying we
get$:$\newline$\dfrac{1}{\upsilon_{k}}\dfrac{\partial u_{k}^{{}}}%
{\partial\beta_{i}}=\dfrac{1}{2\left(  \varepsilon_{k}^{2}+\Delta^{2}\right)
}\left\{  \Delta\dfrac{\partial\varepsilon_{k}}{\partial\beta_{i}}%
-\varepsilon_{k}\dfrac{\partial\Delta}{\partial\beta_{i}}\right\}  $\newline
using $\varepsilon_{k}=\epsilon_{k}-\lambda$, we obtain explicitly:\newline%
$\dfrac{1}{\upsilon_{k}}\frac{\partial u_{k}^{{}}}{\partial\beta_{i}}%
=\dfrac{1}{2\left(  \varepsilon_{k}^{2}+\Delta^{2}\right)  }\left\{
\Delta\dfrac{\partial\epsilon_{k}}{\partial\beta_{i}}-\Delta\dfrac
{\partial\lambda}{\partial\beta_{i}}-\left(  \epsilon_{k}-\lambda\right)
\dfrac{\partial\Delta}{\partial\beta_{i}}\right\}  $\newline Moreover, noting
that:\newline$\dfrac{\partial\epsilon_{k}}{\partial\beta_{i}}=\left\langle
k\right\vert \dfrac{\partial H_{sp}}{\partial\beta_{i}}\left\vert
k\right\rangle $\newline we find:\newline$\dfrac{1}{\upsilon_{k}}%
\dfrac{\partial u_{k}^{{}}}{\partial\beta_{i}}=\dfrac{\Delta}{2\left(
\varepsilon_{k}^{2}+\Delta^{2}\right)  }\left\{  \left\langle k\right\vert
\dfrac{\partial H_{sp}}{\partial\beta_{i}}\left\vert k\right\rangle
-\dfrac{\partial\lambda}{\partial\beta_{i}}-\dfrac{\left(  \epsilon
_{k}-\lambda\right)  }{\Delta}\dfrac{\partial\Delta}{\partial\beta_{i}%
}\right\}  $\newline the quasiparticle energy is $E_{k}=\left(  \varepsilon
_{k}^{2}+\Delta^{2}\right)  ^{1/2}$ so that:\newline$\dfrac{1}{\upsilon_{k}%
}\dfrac{\partial u_{k}^{{}}}{\partial\beta_{i}}=\dfrac{\Delta}{2E_{k}^{2}%
}\left\{  \left\langle k\right\vert \dfrac{\partial H_{sp}}{\partial\beta_{i}%
}\left\vert k\right\rangle -\dfrac{\partial\lambda}{\partial\beta_{i}}%
-\dfrac{\left(  \epsilon_{k}-\lambda\right)  }{\Delta}\dfrac{\partial\Delta
}{\partial\beta_{i}}\right\}  =-\dfrac{\Delta}{2E_{k}^{2}}R_{i}^{k}$\newline
where we have put:\newline%
\begin{equation}
R_{i}^{k}=-\left\langle k\right\vert \dfrac{\partial H_{sp}}{\partial\beta
_{i}}\left\vert k\right\rangle +\dfrac{\partial\lambda}{\partial\beta_{i}%
}+\dfrac{\left(  \epsilon_{k}-\lambda\right)  }{\Delta}\dfrac{\partial\Delta
}{\partial\beta_{i}} \label{rik}%
\end{equation}
\newline Using the result \newline$\dfrac{1}{\upsilon_{k}}\dfrac{\partial
u_{k}^{{}}}{\partial\beta_{i}}=-\dfrac{\Delta}{2E_{k}^{2}}R_{i}^{k}$\newline
the product of the similar terms of Eq. (\ref{prod}) gives finally:\newline%
$\sum_{k}\dfrac{1}{2E_{k}}\dfrac{1}{\upsilon_{k}^{2}}\dfrac{\partial u_{k}%
}{\partial\beta_{i}}\dfrac{\partial u_{k}}{\partial\beta_{j}}=\sum_{k}%
\dfrac{1}{2E_{k}}\left(  \dfrac{1}{\upsilon_{k}}\dfrac{\partial u_{k}%
}{\partial\beta_{i}}\right)  \left(  \dfrac{1}{\upsilon_{k}}\dfrac{\partial
u_{k}}{\partial\beta_{j}}\right)  $\newline$=\sum_{k}\dfrac{1}{2E_{k}}\left(
-\Delta\dfrac{R_{i}^{k}}{2E_{k}^{2}}\right)  \left(  -\Delta\dfrac{R_{j}^{k}%
}{2E_{k}^{2}}\right)  =\sum_{k}\dfrac{\Delta^{2}}{8E_{k}^{5}}R_{i}^{k}%
R_{j}^{k}$\newline The cranking formula of the mass parameters becomes finally:%

\begin{equation}
D_{ij}\left\{  \beta_{1},.,\beta_{n}\right\}  =2\hbar^{2}\sum_{k,l}%
\frac{\left(  u_{k}\upsilon_{l}+u_{l}\upsilon_{k}\right)  ^{2}}{\left(
E_{k}+E_{l}\right)  ^{3}}\left\langle l\right\vert \frac{\partial H_{sp}%
}{\partial\beta_{i}}\left\vert k\right\rangle \left\langle k\right\vert
\frac{\partial H_{sp}}{\partial\beta_{j}}\left\vert l\right\rangle \left(
1-\delta_{k,l}\right)  +2\hbar^{2}\sum_{k}\frac{\Delta^{2}}{8E_{k}^{5}}%
R_{i}^{k}R_{j}^{k} \label{a8}%
\end{equation}


\begin{thebibliography}{99}                                                                                               %


\bibitem {01}A. Arima and F. Iachello, Phys. Rev. Lett. 35, 1069--1072 (1975)

\bibitem {1}M. Brack, J. Damgaard, A. S. Jensen, H. C. Pauli, V. M. Strutinsky
and C. Y. Wong, Rev Mod. Phys. 44 (1972) 320

\bibitem {1a}L. Prochniak, K. Zajac, K. Pomorski, S. G. Rohozinski, J.
Srebrny, Nucl. Phys. A648, 181 (1999)

\bibitem {2}D. R. Inglis, Phys. Rev. 96 (1954) 1059, 97 (1955) 701

\bibitem {2a}A. K. Kerman, Ann. Phys. (New York),12(1961)300, 222, 523

\bibitem {3}M. Baranger, M. V\'{e}n\'{e}roni, Ann. Phys. (NY) 114, 123 (1978).

\bibitem {4}M.J. Giannoni, P. Quentin, Phys. Rev. C21, 2060 (1980).

\bibitem {5}M.J. Giannoni, P. Quentin, Phys. Rev. C21, 2076 (1980).

\bibitem {5a}J. Decharge and D. Gogny, Phys. Rev., C21 (1980) 1568

\bibitem {5b}M. Girod and B. Grammaticos, Phys. Rev., C27 (1983) 2317

\bibitem {5b1}Giraud B., Grammaticos B, Nucl. Phys. A233 , 373, 1974

\bibitem {5bb}M Mirea and R C Bobulescu J. Phys. G: Nucl. Part. Phys. 37
(2010) 055106

\bibitem {5b2}N. Hinohara, T. Nakatsukasa, M. Matsuo and K. Matsuyanagi, Prog.
Theor. Phys. Vol. 115 No. 3 (2006) pp. 567-599

\bibitem {10}J. J. Griffin, Nucl. Phys. A170 (1971) 395

\bibitem {11}T Ledergerber, H. C. Pauli, Nucl. Phys. A207 (1973) 1--32

\bibitem {12}P.-G. Reinhard, Nucl. Phys. A281 (1977) 221--239

\bibitem {14}V. Schneider, J. Maruhn, and W. Greiner, Z. Phys. A 323 (1986) 111

\bibitem {13}D. N. Poenaru, R. A. Gherghescu, W. Greiner, Rom. Journ. Phys.,
Vol. 50, Nos. 1--2, P. 187--197, Bucharest, 2005

\bibitem {8}B. Mohammed-Azizi, and D.E. Medjadi, Computer physics Comm.
156(2004) 241-282.

\bibitem {6}S. T. Belyaev, Mat. Fys. Medd. Dan. Vid. Sehk. 31 (1959) N$%
{{}^\circ}%
$11

\bibitem {7}D. Bes, Mat. Fys. Medd. Dan. Vid. Selsk. 33 (1961) no. 2

\bibitem {5c}E. Kh. Yuldashbaeva, J. Libert, P. Quentin, M. Girod , B. K.
Poluanov, Ukr. J. Phys. 2001. V. 46, N 1
\end{thebibliography}
\end{document}